\documentclass[acmtocl]{acmtrans2m}

\usepackage{amsmath}
\usepackage{graphicx}
\usepackage{algorithm}
\usepackage{subfigure}
\usepackage{epsfig}

\acmVolume{2}
\acmNumber{3}
\acmYear{01}
\acmMonth{09}

\newtheorem{fact}{Fact}
\newtheorem{theorem}{Theorem}
\newcommand{\BibTeX}{{\rm B\kern-.05em{\sc i\kern-.025em b}\kern-.08em
    T\kern-.1667em\lower.7ex\hbox{E}\kern-.125emX}}

\title{On the Effect of Quantum Interaction Distance on Quantum Addition Circuits}
\author
{
BYUNG-SOO CHOI \\ University of Seoul
\and
RODNEY VAN METER \\ Keio University
}

\begin{abstract}
We investigate the theoretical limits of the effect of the quantum interaction distance on the speed of exact quantum addition circuits. For this study, we exploit graph embedding for quantum circuit analysis. We study a logical mapping of qubits and gates of any $\Omega(\log n)$-depth quantum adder circuit for two $n$-qubit registers onto a practical architecture, which limits interaction distance to the nearest neighbors only and supports only one- and two-qubit logical gates. Unfortunately, on the chosen $k$-dimensional practical architecture, we prove that the depth lower bound of any exact quantum addition circuits is no longer $\Omega(\log {n})$, but $\Omega(\sqrt[k]{n})$. This result, the first application of graph embedding to quantum circuits and devices, provides a new tool for compiler development, emphasizes the impact of quantum computer architecture on performance, and acts as a cautionary note when evaluating the time performance of quantum algorithms.
\end{abstract}




\begin{document}

\setcounter{page}{1}

\begin{bottomstuff}
Byung-Soo Choi is with the Center for Quantum Information Processing, Department of Electrical and Computer
Engineering, University of Seoul, Seoul, 130-743, Republic of Korea. (Email:bschoi3@gmail.com. Any correspondence should be sent).
\newline
Rodney Van Meter is with the Faculty of Environment and Information Studies, Keio University, Fujisawa, Japan. (Email: rdv@sfc.wide.ad.jp).
\newline
This paper is an extension of the short abstract which was presented at the Asian Conference on Quantum Information Science, 2008.
\end{bottomstuff}
\maketitle

\section{Introduction}
Quantum computers may outperform classical computers on certain problems. One example is Shor's large number factoring algorithm \cite{shor-factoring}, which can factor a large number within polynomial complexity. For this problem, although there is no proof that polynomial time classical algorithms are impossible, the best known classical algorithm suffers from super-polynomial time complexity \cite{DBLP:conf/crypto/KleinjungAFLTBGKMORTZ10}. As a second example, Grover's database search algorithm \cite{Grover-search} can help to find the desired item from an unstructured database search space of $n$ elements in $O(\sqrt{n})$ computational steps. Since the classical upper bound for the unstructured database search is $O(n)$, the prospect for significant speedup using quantum computers is promising. Many other quantum algorithms have been proposed recently, and the design of quantum algorithms is an active research area \cite{mosca-2008,Bacon-vanDam-2010-Algo-Review}.

In general, the calculation of the speedup of a quantum algorithm over a classical one is based on an ideal quantum computer model, similar to the Random Access Machine model \cite{knuth} for classical computing. Hence, we can say that the quantum speedup in the literature is the ``optimistic'' performance improvement. However, if we take into account the real physical constraints on practical quantum computing machines, the quantum speedup may decrease, especially when considering the circuit depth (time performance). Some upper bounds on the mapping of quantum arithmetic circuits from the ideal model to specific and practical models have been investigated \cite{upper-bound}. On the other hand, very little work has been done to analyze the lower bounds. In this work, we focus on the lower bounds when practical constraints are accounted for the quantum circuit.

While we are interested in the general problem of hardware/software co-design for quantum computers, we focus here on the problem of addition for two $n$-qubit numbers. Addition is a well-defined problem, making direct comparison of competing solutions straightforward. It is also a fundamental building block for important applications, such as Shor's factoring algorithm. For an ideal quantum computer model, such as the arbitrary concurrent (AC) \cite{VBE-Improved} architecture, several types of circuit with a depth of $O(\log n)$ have been designed \cite{draper-QCLA,VBE-Improved}.

In this study, we establish a quantum depth lower bound for adders when the quantum interaction distance is only one, using the nearest-neighbor, two-qubit, and concurrent execution (NTC) \cite{VBE-Improved} architecture. Arithmetic circuits are large and complex unitary transforms, usually decomposed into circuits of one-, two- and three-qubit quantum gates. If the target and source qubits of e.g. a controlled-NOT gate are not neighbors, the target or source qubit must be transported to a neighboring position by using SWAP operations, or using a chain of gates, as shown in Sec.~\ref{sec:ld-gates}. Therefore, on a practical quantum computer, a number of SWAP operations is necessary to emulate the behavior of a quantum adder running on an ideal machine, increasing the circuit depth. In our study, we consider the $k$-dimensional ($k$D) NTC architecture. Because conventional quantum computer architectures are based on one-, two-, and in some cases three-dimensional structures, it might seem to be sufficient to consider only these dimensions for practical use. However, higher dimensions are investigated also for analyzing the general situation; as interconnects improve, we may see structures with higher-dimension interconnects. The Caltech Cosmic Cube \cite{Seitz:1985:CC:2465.2467}, for example, used a six-dimensional hypercube interconnect and others reached ten dimensions \cite{athas:multicomputer} before two- and three-dimensional lattices and tori became the de facto standard for classical multicomputers. Then our question can be rephrased as determining whether or not an $\Omega(\log {n})$-depth quantum adder exists for these models, where $\Omega(f(n))$ indicates that the circuit is asymptotically bounded below by some constant multiple of $f(n)$.

To investigate the theoretical limit of the ideal depth lower bound on these quantum architectures, we exploit some graph theoretical approaches such as \emph{graph embedding}, for the first time. We show that any $\Omega(\log n)$-depth quantum Boolean circuit on the AC architecture can be modeled as a set of log-depth binary trees (LBTs) and that an adder circuit can be described as a set of Boolean circuits, resulting in a set of LBTs where each LBT produces one qubit of the sum. Then the question can be rephrased again to ask how much additional depth is required for embedding a log-depth binary tree into a $k$D graph having edges between neighboring nodes for the corresponding $k$D NTC architecture. In graph embedding for quantum circuits, the additional depth caused by the necessary SWAP operations is measured by the \emph{dilation value}. Based on the analysis of dilation, for embedding a log-depth binary tree into the target graph, we find that the theoretical depth lower bound is $\Omega(\sqrt[k]{n})$ for the $k$D NTC structure. Therefore, there is no $\Omega(\log n)$-depth quantum adder on any $k$D NTC structure by simple logical mapping, due to the practical limitation of the interaction distance.

This work is organized as follows. Section \ref{sec:1} describes several quantum architectures, long-distance quantum gates, quantum Boolean circuits, exact and approximate quantum adders, log-depth binary tree, and graph embedding. Section \ref{sec:2} studies the depth lower bounds of the quantum addition circuits on the target quantum architectures. Section \ref{exam:CLA} describes how a typical $O(\log n)$-depth adder, the carry lookahead adder, can be mapped to a set of log-depth binary trees. Section \ref{sec:related-work} discuss some related work and the differences from ours. Section \ref{sec:3} concludes this manuscript with several research questions.

\section{Background}
\label{sec:1}

\subsection{Quantum Computer Architectures}
Some systems, such as those using ``flying qubits'' held on photons and measurement-based quantum computing \cite{PhysRevA.68.022312}, allow an approximation of arbitrary-distance interaction. At the other extreme, there is the NTC \cite{VBE-Improved} architecture allowing the nearest neighbor interaction only, with one- or two-qubit gates executing concurrently. Since most quantum computer proposals are based on variations of this model, we focus on the NTC model. Depending on the layout of qubits, there are three architectures as follows:

\begin{itemize}
\item {\textbf{1D NTC Model:}} The 1D model, called Linear Nearest Neighbor (LNN) \cite{fowler-2004-4}, consists of qubits located in a single line. In this model, only two neighboring qubits can interact. Some trapped-ion systems~\cite{haeffner05scalable} and liquid nuclear magnetic resonance (NMR) \cite{laforest:012331} technologies are experimental systems based on this model. The original Kane model \cite{kane-model} is also based on this model. The effects of the 1D NTC model on performance have been investigated for the quantum Fourier transform \cite{Takahashi-Improved-Version-of-QFT-for-LNN,VMeter-Topology-Distributed-QFT} and for Shor's algorithm \cite{Shor-1D}.
\item{\textbf{2D NTC Model:}} The 2D NTC model is a lattice structure where the links are located on a two-dimensional Manhattan grid. In this model, a qubit can interact with four neighboring qubits unless, of course, it is on an edge of the grid. Therefore, it can help to reduce the communication cost over the 1D NTC model. Several proposed quantum technologies will correspond to this model, such as the array of trapped ions \cite{haeffner05scalable} and Josephson junctions \cite{citeulike:3014983,PhysRevB.69.214501}.
\item{\textbf{3D NTC Model:}} The 3D NTC model is simply a set of 2D lattices stacked in the third dimension. As expected, since a qubit can interact with six neighboring qubits, it has more flexibility than the 2D NTC model. Although it has some advantages over the 2D NTC model, it suffers from the difficulty of controlling an individual qubit embedded deep in a 3D structure, as well as difficult fabrication. However, some approaches have been proposed based on this model \cite{perez-delgado:100501}.
\end{itemize}

\subsection{Long-Distance Quantum Gates}
\label{sec:ld-gates}

In systems that do not directly support long-distance interactions, we must construct circuits of building blocks using only nearest-neighbor operations. Nearest-neighbor operations can be used in three ways to effect gates between two qubits that are initially stored some distance apart:
\begin{itemize}
\item swap one or more of the qubits we wish to interact along a path in the graph that will bring the qubits together;
\item execute logical gates in a chain along a path so that the end result is the desired gate; or
\item use the graph links to create long-distance entanglement (Bell pairs) that can be used to execute long-distance gates (``telegate'') or to teleport data qubits.
\end{itemize}

We focus primarily on the first method, but let us briefly examine the other two. It is well known that a carefully-chosen chain of neighboring gates can act equivalently to a long-distance gate. For example, on a line of qubits $A,B,C$ with the notation $\operatorname{CNOT}(\textrm{control,target})$ and gates ordered left to right,
\begin{align}
\operatorname{CNOT}(A,C) = \operatorname{CNOT}(A,B) \operatorname{CNOT}(B,C) \operatorname{CNOT}(A,B) \operatorname{CNOT}(B,C).
\end{align}
This approach results in identical asymptotic circuit depth and complexity as the swapping approach, as the gates must be cascaded in an identical fashion.  Constant factors can vary, however, as a result of the usage pattern of the variables and the gate execution time; in general, the principle of locality \cite{hennessy-patterson:arch-quant4ed} suggests moving the variable will be more effective than using the gate chain method.

A long-distance Bell pair can be created using pairwise entangling gates along a path in the connectivity graph, measuring the middle qubits, and propagating a Pauli frame correction to the end points, as done in quantum repeaters and measurement-based quantum computation \cite{dur:PhysRevA.59.169,PhysRevA.68.022312}. The quantum operations in this approach can be executed in only two time steps; however, the classical information will be limited by the speed of signal propagation in the system. This limitation assumes that non-Clifford group operations are executed at each end of the movement. For our purpose, this restriction holds, as addition circuits require non-Clifford group operations. Equally important, this approach consumes significant spatial resources: the intermediate qubits along the path cannot hold important data values, as they are measured and discarded.

Browne et al. have recently shown that one-way quantum computation (measurement-based quantum computation) is equivalent in power to unbounded-fanout circuits \cite{browne-equivalence-MBQC-CBQC-with-fanout}. Our results are argued using both the fanout and the computational aspects of the problem.

Thus, the results presented here are restricted: they are not yet shown to apply to measurement-based quantum computation, they assume that classical signal propagation is restricted to the same connectivity as the quantum operations, and the operations of interest before and after data movement must be non-Clifford group operations. However, preparing large cluster states for the measurement-based quantum computation is still very hard \cite{PhysRevA.81.042322,PhysRevA.81.052301,PhysRevA.82.022331}, and our results apply to many of the most promising quantum computer architectures, including those based on Calderbank-Shor-Steane codes \cite{Shor-9-Code,PhysRevA.54.1098,Steane-7-Code,PhysRevLett.77.793,rod-thesis}.

\subsection{Quantum Boolean Circuit}

A classical Boolean circuit is a circuit for $n$ inputs with one output. Since the number of outputs is one and the value of the output is zero or one, sometimes a classical Boolean circuit can be called a binary decision circuit. As with classical Boolean circuits, in a quantum Boolean circuit, our goal is to compute a single output qubit that is a function of the $n$ input qubits. The final output is stored in the output qubit, and any ancillae may be cleaned by undoing the computation.

\subsection{Quantum Addition}

In-place addition on a quantum computer performs the transform $|a,b\rangle \rightarrow |a,a+b\rangle$, where $|a\rangle$ and $|b\rangle$ are $n$-qubit registers holding binary numbers. If we consider each summation output as a single output qubit, the quantum addition circuit consists of a set of quantum Boolean circuits, one for each output qubit.

Numerous quantum addition algorithms have been proposed, and even implemented at small scales, based on classical addition algorithms. Ripple-carry algorithms include those proposed by Vedral et al. \cite{VBE-adder}, Beckman et al. \cite{BCDP-adder}, Cuccaro et al. \cite{CDKM-adder}, and Takahashi \cite{takahashi-survey}. The depth of ripple-carry adders is linear in the length of the numbers being added, and they typically do not require long-distance interactions. Logarithmic-depth adders, including the carry-lookahead and conditional-sum adders, have been designed using longer-distance operations \cite{draper-QCLA,VBE-Improved}, assuming the AC abstraction architecture; one of these has been adapted to measurement-based quantum computation \cite{trisetyarso-2009}.

Note that the above adders are exact, rather than approximate, integer adders. That is, we expect that $|0111...11\rangle+|0000...01\rangle$ will yield the result $|1000...00\rangle$. However, we can consider non-exact adders.  Draper proposed an $O(\log n)$-depth adder based on the quantum Fourier transform \cite{Draper-QAddition}. This adder is quite different from the above adders since it is based on a genuine quantum approach, rather than classical techniques. While for approximate arithmetic the circuit depth may be less than $O(n)$, in order to achieve full $n$-qubit precision, the depth is $O(n)$.

\subsection{Log-depth Binary Tree}

\begin{figure*}
\centering
\includegraphics[scale=0.8]{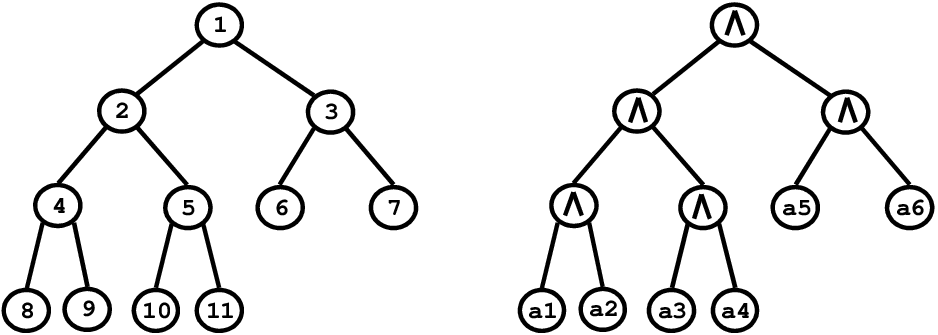}
\caption{An example of a log-depth binary tree and its application. \newline
(\textbf{left}) The root node 1 has two children, 2 (left child) and 3 (right child). Likewise, non-leaf nodes 2 and 3 each have two children. The smallest depth of the log-depth binary tree is $\lceil \log{n} \rceil$, where $n$ is the number of nodes. \newline
(\textbf{right}) A mapping of a 6-bit AND circuit $a_1 \wedge a_2 \wedge a_3 \wedge a_4 \wedge a_5 \wedge a_6$ with two-input AND gate is shown. A circuit with two-input gates can be modeled as a log-depth binary tree where the initial input is mapped to the leaf nodes, all two-input gates to the non-leaf nodes, and the final output to the root-node. }
\label{log-depth-binary-tree}
\end{figure*}

A log-depth binary tree is defined as a class of binary tree \cite{LBT} that has one root, one or two child nodes from each non-leaf node, and all other leaf nodes. In a tree, the depth can be defined as the number of nodes in the longest path from the root to any leaf node. For a log-depth tree, the highest depth must be $O(\log n)$, when the number of leaves is $n$. Figure \ref{log-depth-binary-tree} is an example of a log-depth binary tree and its application. Since many digital algorithms are based on binary decisions with one- or two-input gates, the log-depth binary tree is a very useful model. Likewise, many acyclic circuits with one- or two-input gates can be modeled as log-depth binary trees. Therefore, we use the log-depth binary tree for the analysis of arithmetic quantum circuits on the NTC architecture.

\subsection{Graph Embedding}
\label{sec:graph-embedding}

\begin{figure*}
\centering
\includegraphics[scale=0.8]{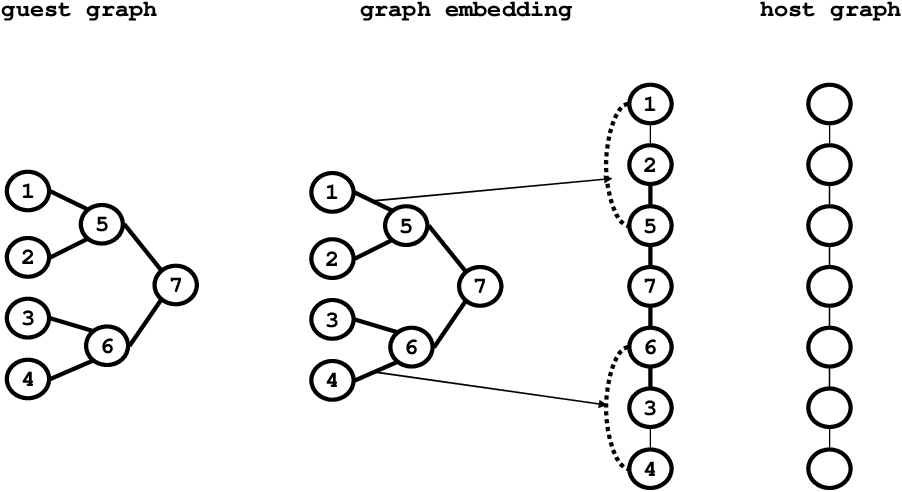}
\caption{Embedding a \textbf{log-depth binary tree (leftmost)} into a \textbf{line graph (rightmost)}. \newline Dotted lines represent the graph dilation. The edges (1,5) and (4,6) in the log-depth binary tree cannot be directly mapped to any edges in the line graph. They are mapped to paths (1,2)\&(2,5) and (4,3)\&(3,6), respectively.}
\label{graph-embedding}
\end{figure*}

Graph embedding is a widely used tool for analyzing the performance of different structures (see e.g. Diestel \cite{Book-Graph-Theory}). A guest graph $G$ is embedded on a host graph $H$ when the nodes in $G$ are mapped to the nodes in $H$, and the edges in $G$ are mapped to paths in $H$. Figure \ref{graph-embedding} shows an example of embedding a log-depth binary tree (leftmost) into a line graph (rightmost). Each node in $G$ is mapped to a node in $H$. The two edges (1,5) and (4,6) in $G$ cannot be directly mapped to any edge in $H$, but can be mapped to paths, as shown by the dotted lines. Graph embedding has many interesting properties \cite{651501}:

\begin{itemize}
\item   {\textbf{dilation:}} The dilation is defined as the maximum distance between adjacent nodes in $H$ after embedding. In general, the dilation lower bound is calculated as \cite{unger}
    \begin{equation}
    \frac{diameter\,\, of\,\, the\,\, host\,\, graph}{diameter\,\, of\,\, the\,\, guest\,\, graph}.
    \label{diameter-equation}
    \end{equation}

    In the above equation, the diameter of a graph $G$ is defined as the maximum distance between any two nodes in $G$, where the distance between two nodes $u$ and $v$ is the number of hops in the shortest path between $u$ and $v$. The lower bound of dilation occurs when there is a mapping with the smallest increase of the distance between nodes in the guest graph. To achieve this, first, we can map the longest path (diameter) of the guest graph to the longest path of the host graph. Then, for the remaining paths, we choose the longest path again and map it to the longest path of the host graph. We repeat this procedure until all paths find the corresponding paths in the host graph. During each round, the path of the guest graph has to be mapped to a path of the host graph with the same or larger distance. Therefore, the lowest ratio of the diameters for the guest and the host graphs is the lower bound of the dilation value. For example, as shown in Figure \ref{graph-embedding}, the dilation is two since the edge (1,5) in $G$ must be embedded into a path (1,2)\&(2,5) in $H$. Therefore, to emulate the interaction between 1 and 5 in $G$, two interactions are required, between 1 and 2, and 2 and 5, in $H$.
\item   {\textbf{expansion:}} The expansion is defined as the ratio of the number of nodes in $H$ over the number of nodes in $G$.
\item   {\textbf{load:}} The load is defined as the maximum number of nodes in $G$ which must be embedded into a node in $H$.
\item   {\textbf{congestion:}} The congestion is defined as the maximum number of edges in $G$ which must be embedded into an edge in $H$.
\end{itemize}

In our study we consider only the dilation. In circuit complexity studies \cite{vollmer1999introduction}, the dilation can be used to find the depth increase of a base circuit when the circuit has to be mapped to a certain architecture.

\section{Depth Lower Bounds}
\label{sec:2}

First, we need to understand the depth lower bound of quantum Boolean circuits for $n$ inputs when only one- and two-qubit gates are allowed, with no limitation of interaction distance.

\begin{fact}
\label{log-depth-of-exact-boolean-function}
The depth lower bound of an exact quantum Boolean circuit for $n$ inputs is $\Omega(\log n)$ when only one- or two-qubit gates are allowed, without limitation of interaction distance.
\end{fact}

\begin{proof}
We consider the general structure of any $\Omega(\log {n})$-depth quantum Boolean circuit. Any $\Omega(\log {n})$-depth quantum Boolean circuit can be executed in the following manner. For generating the final output qubit, a two-qubit gate (which we will place at the root of the binary tree) must be applied to two temporary input qubits which are generated at the previous level. These temporary two input qubits are generated from other two-qubit gates with each set of temporary two input qubits which are generated at the previous level. This backtracking must continue until the temporary input qubits are the same as the actual input qubits. We need to calculate how many levels are needed. Since each two-qubit gate needs two inputs, the number of temporary inputs doubles. Hence, if the level of backtracking is $k$, then the number of inputs is $2^k$. Therefore, the minimum level of backtracking or levels must satisfy $2^k \geq n$, and hence $k \geq \lceil \log n \rceil $. In this manner, we can make any $\Omega(\log n)$-depth quantum Boolean circuit with one- and two-qubit gates, as explained by Cleve and Watrous \cite[P.532]{Cleve-FastParallelCircuitForQFT}.
\end{proof}

Next, we need to investigate the graph structure of an $\Omega(\log {n})$-depth quantum Boolean circuit.

\begin{theorem}
\label{log-depth-boolean-circuit-to-log-depth-tree}
Any $\Omega(\log {n})$-depth quantum Boolean circuit can be represented by a log-depth binary tree when only one- or two-qubit gates are allowable with no limitation of interaction distance.
\end{theorem}

\begin{proof}
The one- and two-qubit gates in a quantum Boolean circuit can be mapped to non-leaf nodes and the root node in the log-depth binary tree. The root node contains the final output. The actual inputs can be mapped to leaf nodes. The two inputs for each two-qubit gate can be mapped to the left and the right child nodes for the corresponding parent node. In this manner, we can map any $\Omega(\log {n})$-depth quantum Boolean circuit into a log-depth binary tree. Note that the edges in the graph represent the information flow from the child node to the parent node. The time for communication of information between the child and the parent node is ignored in this analysis.
\end{proof}

\begin{figure*}
\centering
\includegraphics[scale=0.8]{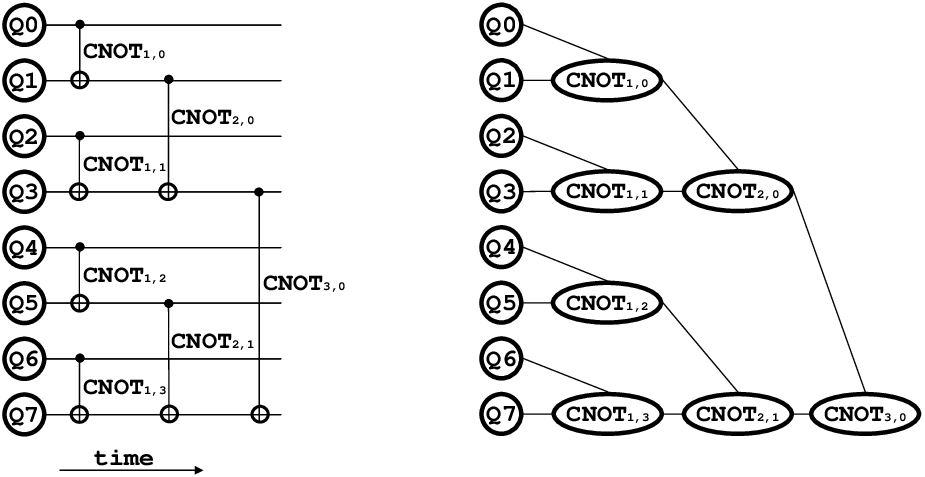}
\caption{Mapping an \textbf{$\Omega(\log {n})$-depth quantum Boolean circuit (left)} for an 8-bit PARITY function into a \textbf{log-depth binary tree (right)}. \newline The inputs of the left are mapped to the leaf nodes in the right. Two-qubit gates in the left are mapped to non-leaf nodes in the right. Final output is generated on the root node}
\label{PARITY-in-LBT}
\end{figure*}

As an example, a mapping of a quantum Boolean circuit for an 8-qubit PARITY function into a log-depth binary tree is shown in Figure \ref{PARITY-in-LBT}. In the first level, four CNOT operations -- $CNOT_{1,0}$, $CNOT_{1,1}$, $CNOT_{1,2}$, and $CNOT_{1,3}$ -- are applied to the corresponding qubits. The outputs are stored in $Q_1$, $Q_3$, $Q_5$, and $Q_7$, respectively. In the second level, two CNOT operations -- $CNOT_{2,0}$ and $CNOT_{2,1}$ -- are applied to the output qubits from the first level. The results are stored in $Q_3$ and $Q_7$. In the last level, one CNOT operation $CNOT_{3,0}$ is applied, and the result is stored in $Q_7$. Now we can map this circuit into a log-depth binary tree, as shown in the right part of Figure \ref{PARITY-in-LBT}. In the figure, input qubits are mapped to the leaf nodes. The CNOT operations in the circuit are mapped to the non-leaf nodes in the log-depth binary tree. The final output is stored in the root node.

\begin{theorem}
\label{sum-lbt}
A quantum Boolean circuit for a summation output $s_i$ can be mapped to a log-depth binary tree when only one- and two-qubit gates are used without limitation of interaction distance.
\end{theorem}

\begin{proof}
A summation output $s_i$ can be generated by an exact quantum Boolean circuit for $n$ inputs since the input carry for $s_i$ position depends on all $a_i$ and $b_i$ where $i \in \{0,\cdots, i-1\}$. Therefore, the depth lower bound of the quantum Boolean circuit for $s_i$ is $\Omega(\log n)$ by Fact \ref{log-depth-of-exact-boolean-function}. Since a quantum Boolean circuit with $\Omega(\log n)$-depth can be mapped to a log-depth binary tree as shown by Theorem \ref{log-depth-boolean-circuit-to-log-depth-tree}, a quantum Boolean circuit for a summation output $s_i$ can be mapped to a log-depth binary tree.
\end{proof}

Up to this point, we have discussed a quantum Boolean circuit for $s_i$ and its log-depth binary tree structure. To reduce the overall addition time, each summation output $s_i$ must be generated as fast as possible, so the quantum Boolean circuits for all output qubits $s_i$ must be executed in parallel. However, since each summation output $s_i$ needs to use the inputs $a_j$ and $b_j$, where $j \in \{0, \cdots, i\}$, copies of the inputs $a_j$ and $b_j$ must be prepared for each quantum Boolean circuit for $s_i$. Therefore, each input $a_j$ and $b_j$ must be fanned out for each quantum Boolean circuit for $s_i$, and hence must be fanned out at most $n-j$ times.

\begin{fact}
\label{dup-lbt}
A fanout circuit for a single qubit to $n$ target qubits can be mapped to a log-depth binary tree.
\end{fact}

\begin{figure*}
\centering
\includegraphics[scale=0.6]{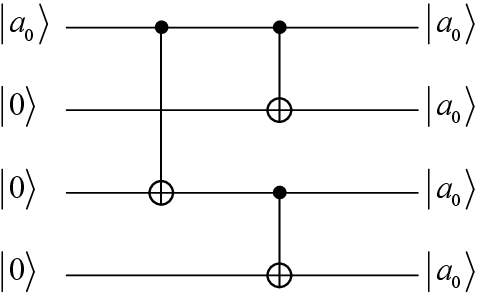}
\caption{An input qubit $|a_0\rangle$ is fanned out four times in the quantum equivalent of a FANOUT circuit. The depth of the circuit is $\lceil \log{n} \rceil$, where the number of fanout outputs is $n$.}
\label{input-qubit-fanout}
\end{figure*}

\begin{proof}
We prove this by construction. For example, an input qubit $|a_0\rangle$ can be fanned out four times as shown in Figure~\ref{input-qubit-fanout}. Since a CNOT gate fans out one input into two outputs, the depth lower bound is $\Omega(\log n)$.
\end{proof}

Now, we want to know the depth lower bound of any quantum addition circuit when only one- and two-qubit gates are allowed with no limitation of interaction distance. For the general case, we define the gate width $g$ to mean that one- to $g$-qubit gates are allowed.

\begin{theorem}
\label{depth-optimal-adder}
On a quantum computer architecture of limited gate width $g$, no quantum adder can be asymptotically faster than one composed of a set of quantum Boolean circuits, where each circuit can be mapped to a log-depth binary tree.
\end{theorem}

\begin{proof}
This theorem is founded on the information flow in an addition circuit.

First, we discuss the case of the gate width $g=2$, meaning only one- and two-qubit gates are allowed. To minimize the depth of the circuit, we need to maximize the parallelism in the circuit. Hence, we can consider the overall circuit to consist of $n+1$ separate quantum Boolean circuits, one to calculate each output qubit (including the final carry out). We call this phase of the circuit the computation part.

The gate width limits the use of each input qubit. Because the $i$-th output qubit depends on all of the input qubits $|a_j\rangle$ and $|b_j\rangle$ for all $j \le i$, each quantum Boolean circuit for each output must have its own copy of the input qubits before execution, in order to run concurrently. Therefore, we need to construct another circuit to fanout the input qubits, making one copy for each tree\footnote{Note this is not quantum cloning, but quantum fanout by using CNOT gates.}. We call this phase the fanout part of the addition circuit.

After the fanout of the input qubits, the $n$ summation and one output carry quantum Boolean circuits are executed in parallel.

Similar arguments follow for any fixed gate width $g > 2$.

Now we consider a graph structure for quantum addition circuit. As we already discussed, the addition consists of two parts: the fanout part and the computation part.  By Fact \ref{dup-lbt} and Theorem \ref{sum-lbt}, a quantum addition circuit can be mapped to a set of log-depth binary trees.
\end{proof}

We observe that this construction is efficient in time, but not in space; $O(n^2)$ physical qubits are required. In practice, both the carry-lookahead and conditional-sum adders (the two known types of $O(\log {n})$-depth quantum adders) do not require the full fanout of data, but reuse the input qubits and partial results more efficiently.  However, the proof above shows that no circuit can do {\em better} than this construction in the circuit depth.

We have now explained how to map a quantum addition circuit to a set of log-depth binary trees. Next, we show how to embed such a log-depth binary tree into a kD mesh structure, which is the graph structure for the kD NTC architecture.

\begin{fact}
\label{log-kd-embedding}
A log-depth binary tree can be mapped to a $k$D mesh with dilation $\Omega(\frac{\sqrt[k]{n}}{\log n})$, hence the depth lower bound of the embedded graph is $\Omega(\sqrt[k]{n})$.
\end{fact}

\begin{proof}
To understand the effect of graph embedding, we need to calculate the dilation of embedding a guest graph into a host graph. The dilation of a graph mapping is the ratio of the diameters. Formally, the dilation for graph mapping a guest graph to a host graph is calculated by Equation (\ref{diameter-equation}) in Section \ref{sec:graph-embedding}.

In our study, the guest graph is a log-depth binary tree whose diameter is $\Omega(\log {n})$ since the max distance is between the two leaves where the path passes through the root node. On the other hand, the host graph is a $k$D mesh graph whose diameter is $\Omega(\sqrt[k]{n})$. Therefore, the dilation of graph embedding from a log-depth binary tree into a $k$D mesh graph is $\Omega(\frac{\sqrt[k]{n}}{\log {n}})$ as shown by Heckmann et al. \cite{Heckmann91optimalembedding}. Finally, the depth of the embedded graph is $\Omega(\frac{ \sqrt[k] {n} }{ \log {n} }) * \Omega(\log {n})  = \Omega( \sqrt[k] {n})$, since the depth of the guest graph increases by the dilation factor.
\end{proof}

\begin{theorem}
The depth lower bound of the exact quantum addition circuit on the $k$D NTC structure is $\Omega(\sqrt[k]{n})$.
\end{theorem}

\begin{proof}
By Theorem \ref{depth-optimal-adder}, a depth-optimal quantum adder can be mapped to a set of log-depth binary trees. By Fact \ref{log-kd-embedding}, a log-depth binary tree can be embedded in a $k$D mesh with a depth of $\Omega( \sqrt[k] {n})$. Therefore, a depth-optimal exact quantum addition circuit in the AC architecture can be mapped into the $k$D NTC architecture with a depth of $\Omega(\sqrt[k]{n})$.
\end{proof}

Therefore, there is no $\Theta(\log {n})$-depth quantum adder on the $k$D NTC quantum computer model.

\section{Case Study: Carry Lookahead Adder}
\label{exam:CLA}

\begin{figure*}
\centering
\includegraphics[scale=0.8]{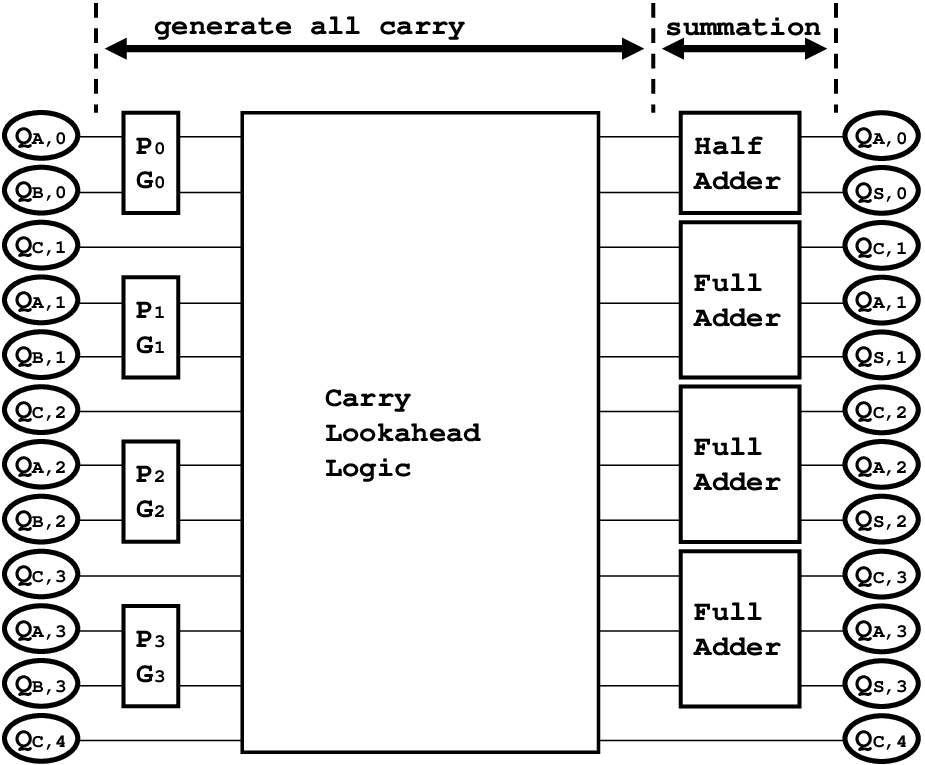}
\caption{Two steps of addition for a $\Omega(\log n)$-depth adder for $n$ qubits. \newline In the first step, carry values for each position are generated concurrently. This step is $\log n$ depth. In the second step, each summation output is generated concurrently by using the corresponding carry value.}
\label{two-step}
\end{figure*}

We show how a carry-lookahead adder (CLA) can be mapped to a set of log-depth binary trees as follows. Let us consider the computation part first. Conceptually, the computation part of the CLA works in two steps: 1) find all $i$-th carry value concurrently and then 2) generate $i$-th summation value concurrently, as shown in Figure \ref{two-step}.

A carry-lookahead adder consists of three networks: \emph{generate} for $g_i$, \emph{propagate} for $p_i$, and \emph{carry-lookahead} for $c_i$. The final results $s_i=a_i \oplus b_i \oplus c_i$ will be calculated for all $i$ in parallel.

In the first step, $g_i=a_i \wedge b_i$ and $p_i=a_i \oplus b_i$ are generated at the same time for all $i$. Since $g_i$ and $p_i$ depend only on $a_i$ and $b_i$, there is no information dependency and hence each $g_i$ and $p_i$ can be generated concurrently. On the other hand, the carry $c_{i+1}$ is generated by using carry lookahead logic
\cite{digital-arithmetic} as follows.
\begin{eqnarray}
\label{carry-equation}
c_{i+1}     &   =   &   g_i +   p_i \wedge c_i\\ \nonumber
            &   =   &   g_i +   p_i \wedge(g_{i-1}+p_{i-1} \wedge c_{i-1})\\ \nonumber
            &   =   &   g_i +   p_i \wedge(g_{i-1}+p_{i-1} \wedge (g_{i-2}+p_{i-2} \wedge c_{i-2}))\\ \nonumber
            &   \vdots  & \nonumber \\ \nonumber
            &   =   &   g_i +   p_i \wedge(g_{i-1}+p_{i-1} \wedge (g_{i-2}+p_{i-2} \wedge (\cdots(g_0+p_0 \wedge c_0)\cdots)))\\ \nonumber
            &   =   &   g_i +   p_i \wedge g_{i-1}+p_i \wedge p_{i-1} \wedge g_{i-2}+\cdots+p_i \wedge p_{i-1} \wedge p_{i-2} \wedge \cdots \wedge p_0 \wedge c_0. \nonumber
\end{eqnarray}
As the above equation explains, $c_{i+1}$ depends on all $g_j$ and $p_j$ where $j \in \{0, \cdots, i\}$.

Although the final summation $s_i=a_i \oplus b_i \oplus c_i$ depends on $a_i$, $b_i$, and $c_i$, the depth is bounded by the circuit for $c_i$. From Equation (\ref{carry-equation}), we know that the carry-lookahead logic consists of the summation of products. Therefore, in the first step, each product term must be generated, and then all products must be summed. As a result, we need to map each product into a log-depth binary tree, and the last summation part into another log-depth binary tree.

\begin{figure*}
\centering
\subfigure[\label{LDB-form-for-AND}]{\epsfig{file=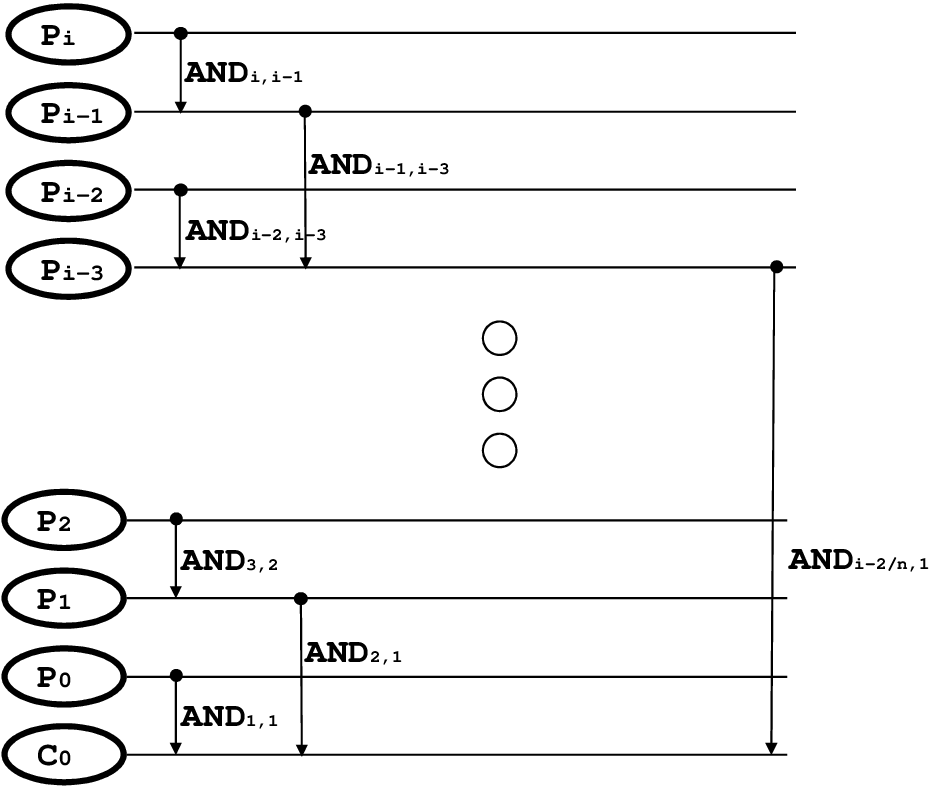,scale=0.5}}
\subfigure[\label{CCNOT-SWAP-for-AND}]{\epsfig{file=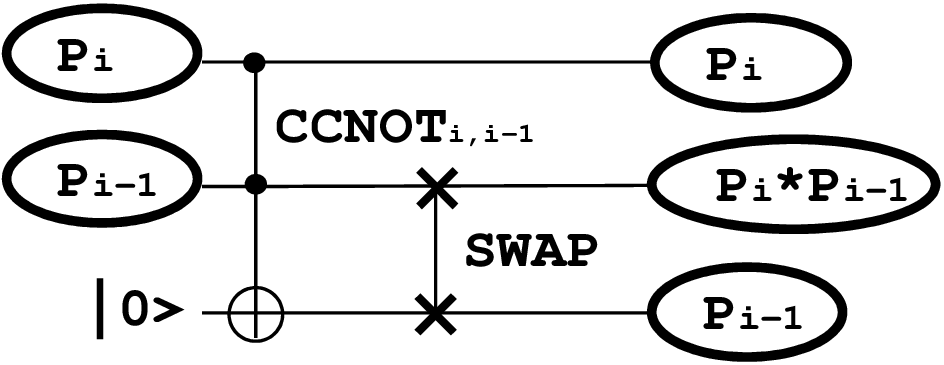,scale=0.3}}
\subfigure[\label{ONE-TWO-SWAP-gates-for-AND}]{\epsfig{file=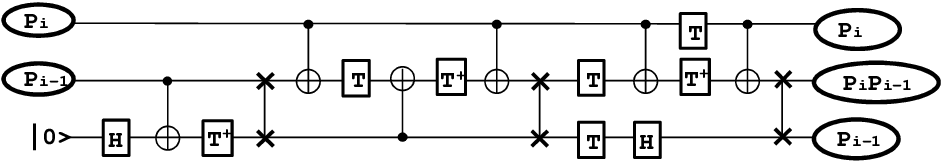,scale=0.7}}
\caption{Two-qubit gate implementation of an AND log-depth binary tree.
\newline (\textbf{a}) A log-depth binary form for $(i+2)$-input AND function.
\newline (\textbf{b}) A decomposition of two-input AND gate with CCNOT and SWAP gate.
\newline (\textbf{c}) A decomposition of two-input AND gate with one-input, two-input, and SWAP gates for satisfying the NTC architecture constraints. The given decomposition is based on Figure 1 of \protect\cite{CNOTs-for-Toffoli-gate} with Hadamard gate $H$, $T = $exp$(i \pi s_z/8)$, and its conjugate $T^\dagger$, which are suitable for fault-tolerant gate implementation.}
\label{test}
\end{figure*}

Let us first consider the product terms. Although there are many products, it is sufficient to consider the worst case $p_i \wedge p_{i-1} \wedge p_{i-2} \wedge \cdots \wedge p_0 \wedge c_0$ since other products can be mapped in the same way. This product is generated as the AND function of $i+1$ $p_i$ values and $c_0$. An AND function for $i+2$ inputs can be implemented by using a log-depth binary tree with some additional qubits as shown in Figure \ref{LDB-form-for-AND}.

Since a two-qubit AND gate cannot be implemented directly, we use CCNOT and SWAP gates for it as shown in Figure \ref{CCNOT-SWAP-for-AND}. Note that this construction needs one ancilla since the two-input AND gate cannot be designed as a unitary gate without using an extra qubit, which increases the overhead. However, this overhead is linear in this case since the maximum overhead is $n$ because the number of AND gates is $n$. Although the SWAP operator is not technically necessary, we introduce it in order to have a consistent representation, storing the output on one of the input qubits.

Because the NTC architecture allows only one- or two-qubit gates, we must further decompose the CCNOT gate; one such decomposition is shown in Figure \ref{ONE-TWO-SWAP-gates-for-AND}. The given decomposition is based on Figure 1 of \protect\cite{CNOTs-for-Toffoli-gate} with Hadamard gate $H$, the gate $T = $exp$(i \pi s_z/8)$, and its conjugate $T^\dagger$, which are suitable for fault-tolerant gate implementation. Therefore, each AND gate in the log-depth binary tree can be implemented by a constant number of one- and two-qubit gates, which increases the coefficient part of the circuit complexity. Note that the ancilla can be initialized again after completing the whole addition by uncomputing in the usual fashion \cite{bennet1973}. In this way, we can generate each log-depth binary tree for each product.

Now let us consider the final summation of products. The summation in the Boolean function requires an OR function, and the structure of the OR function is almost same as the AND function. Hence, we can generate another log-depth binary tree for this summation circuit in the same way.

Thus, we can find a log-depth binary tree for generating the $i$-th carry value. Then, the final output can be generated by using this value with other $a_i$ and $b_i$ values. In this manner, we finally can find a set of log-depth binary trees for each output.

As we discussed in the previous section, the CLA needs another circuit for fanout of inputs for parallel computation for $c_i$. The necessary log-depth binary tree can be built using the fanout circuit shown in Figure \ref{input-qubit-fanout}.

Therefore, for fanout of inputs and for computation of outputs, we can find a set of log-depth binary trees.

\section{Related Work}
\label{sec:related-work}

The effects of the NTC constraints on quantum computation have been studied by several groups.

First, the effect of interaction distance on the quantum circuit synthesis has been investigated. M\"{o}tt\"{o}nen and Vartiainen studied the decomposition of a uniformly controlled gate into one-qubit and CNOT gates \cite{Mottonen-Decomposition-QGates}. They also investigated the effect of interaction distance on the control and target qubits on the nearest neighbor architecture. They shown that the number of gates for one-qubit and CNOTs does not dramatically increase. Similarly, Shende et al. \cite{Shende-SynthesisQLogicCircuits} also studied the synthesis of quantum-logic circuits. They proposed quantum multiplexor circuits, which are elementary circuits for synthesizing a given $n$-qubit circuit. They investigated the overhead when the architecture is limited to a linear nearest neighbor architecture, and showed that the LNN architecture increases the depth by a constant factor of nine times over the generic case. Note that the limitation of interaction distance causes some overhead in the number of gates because they focused on the general case. In this paper, the focus is on the special case of addition.

Second, several quantum circuits have been redesigned with NTC constraints. Maslov investigated circuits for the quantum Fourier transform and the stabilizer code in the LNN architecture \cite{Maslov-Stabilizer-FT}, as did Takahashi \cite{Takahashi-Improved-Version-of-QFT-for-LNN}. Maslov showed that these circuits can be mapped to an LNN architecture with linear depth limited because of interaction distance. Maslov et al. \cite{Maslov-QCircuitPlacement} investigated the technical mapping of logical qubits to the physical qubits. Since the mapping of logical qubits to the physical qubits affects the quantum gate time, very similar to interaction distance, they show that the overall computation time heavily depends on the qubit mapping, which they call the quantum circuit placement problem. Fowler et al.  \cite{fowler-2004-4} investigated how to efficiently map Shor's algorithm on the 1D NTC architecture.

Third, the effect of interaction distance on the error threshold value has been investigated. Gottesman showed the existence of the error thresholds for 1D-, 2D-, and 3D-NTC architectures \cite{exist-pth-NTC-architecture}. Szkopek et al. \cite{p-th-on-NTC} have investigated the effect of the limited interaction distance on the CSS code \cite{PhysRevA.54.1098,PhysRevLett.77.793} threshold values. Based on their work, many SWAP gates are necessary for implementing long distance interaction on the NTC architecture, causing a factor of about 175 times penalty in error threshold. Svore et al. \cite{PhysRevA.72.022317} also investigated the locality issue on the fault-tolerance threshold for quantum computation. Specifically, they showed that the error threshold decreases with the spatial scale-up because of coding in the NTC architecture. In general, since the NTC architecture needs additional SWAP operations for implementing long distance two-qubit gate operations, increasing the number of cases of logical errors, the error threshold value decreases with NTC constraints. Because the logical and physical topologies presented are separate, these results have no direct bearing on our arithmetic results presented here, but are indicative of the importance of the topology. Application of our graph embedding methods to these circuits would be valuable future work.

\section{Conclusion and Open Problems}
\label{sec:3}

We have investigated the effect of the allowed quantum interaction distance on the performance of arithmetic circuits. Since the proposed quantum addition circuits, such as the carry-lookahead adder, were defined with no limitation on the allowed quantum interaction distance, the depth lower bound shown in some previous papers is near to the ideal limit of $O(\log {n})$. However, as we have shown in this work, when the quantum interaction distance is one, the quantum addition circuit must use a number of SWAP operations. Unfortunately, some of the SWAP operations will be in the longest path in the circuit, and hence will increase the depth lower bound.  While this restriction has been recognized in practical terms in some other papers \cite{fowler-2004-4,VBE-Improved,Shor-1D}, it has not had a formal basis. In this study, we investigated a logical mapping of adders defined on the AC architecture into adders for the $k$D NTC architecture, showing $\Omega(\sqrt[k]{n})$ depth because of a practical limitation, the interaction distance by exploiting graph embedding, for the first time. Therefore, we can conclude that when the interaction distance is limited to one, there is no $\Omega(\log {n})$ depth exact quantum addition circuit on any $k$D NTC structure by using simple logical mapping.

We should note that these results apply to the \emph{logical} structure of the systems; the \emph{physical} structure may differ due to the impact of quantum error correction on the physical arrangement of qubits. Also, our method can be applied to analyze reversible classical circuits as well as quantum circuits.

Although the exact quantum integer adder circuit is an important circuit, it is also desirable to analyze other quantum arithmetic circuits in the same fashion. For example, it would be interesting to investigate multipliers, modulo adders, and multipliers over ${Z}_p$ or $GF(2^n)$, as well as other application circuits.

An important future extension of this work is to apply these techniques to the measurement-based quantum computation \cite{1WQC-origin,PhysRevA.68.022312} model or an architecture with topological error correction such as the surface code \cite{TCSQC-RHK,RevModPhys.80.1083,PhysRevLett.101.010501,fowler-2009-9,vanmeter-2009,cody-layered-architecture}. We also intend to expand our results to more general one- and two-qubit unitaries. Finally, as noted in the last section, applying these techniques to error correction circuits may prove enlightening.

The only tool from graph theory that we have used in this study is the dilation property of graph embedding. However, graph embedding has many other interesting properties which may affect the layout of final quantum arithmetic circuit on a specific graph structure. For example, the \emph{congestion}, \emph{expansion}, and \emph{load} are also important  \cite{651501}, and their effects on quantum arithmetic circuits for the 1D, 2D, and 3D structures should be studied. We may investigate the results of Bein et. al. \cite{10.1109-ISPAN.2000.900278} in the view of embedding quantum arithmetic circuits in $k$D NTC structures.

\begin{acks}
This work was supported by the National Research Foundation of Korea Grant funded by the Korean Government (Ministry of Education, Science and Technology).[NRF-2010-359-D00012]
\end{acks}

\bibliographystyle{acmtrans}


\begin{received}
Received February ???;
November ???;
accepted January ???
\end{received}

\end{document}